\documentclass[aps,prd,twocolumn,preprintnumbers,superscriptaddress,
               nofootinbib,floatfix,showkeys]{revtex4-2}

\usepackage{amsmath,amssymb,amsfonts}
\usepackage{graphicx}
\usepackage{dcolumn}
\usepackage{bm}
\usepackage{hyperref}
\usepackage{xcolor}
\usepackage{booktabs}
\usepackage{multirow}
\usepackage{float}
\usepackage{tikz}
\usepackage{pgfplots}
\usepackage{lineno}
\pgfplotsset{compat=1.18}
\usetikzlibrary{arrows.meta, positioning, shapes.geometric,
                decorations.pathmorphing, decorations.markings,
                calc, patterns, backgrounds}

\definecolor{srcblue}{RGB}{31,119,180}
\definecolor{mecorange}{RGB}{214,107,0}
\definecolor{fsigreen}{RGB}{44,160,44}
\definecolor{darkgray}{RGB}{80,80,80}

\newcommand{\prel}{p_{\mathrm{rel}}}
\newcommand{\pcm}{p_{\mathrm{CM}}}
\newcommand{\krel}{k_{\mathrm{rel}}}
\newcommand{\kcm}{k_{\mathrm{CM}}}
\newcommand{\RR}{R_{\mathrm{NC/CC}}}
\newcommand{\DR}{\mathcal{D}_{\mathrm{NC/CC}}}
\newcommand{\numubar}{\bar\nu_\mu}

\tikzset{
  boson/.style={decorate,
    decoration={snake, amplitude=2pt, segment length=6pt,
                post length=2pt, pre length=2pt}},
  fermion/.style={postaction={decorate},
    decoration={markings, mark=at position 0.6 with {\arrow{Latex}}}},
  antifermion/.style={postaction={decorate},
    decoration={markings, mark=at position 0.6 with {\arrowreversed{Latex}}}},
  nucleus/.style={circle, draw=darkgray, fill=darkgray!10,
                  minimum size=22pt, font=\scriptsize},
  vertex/.style={circle, draw=black, fill=black, minimum size=4pt,
                 inner sep=0pt},
}

\begin{document}

\title{A Differential Neutral-to-Charged Current Double-Proton Observable for
Studying Short-Range Correlations in Neutrino--Argon Scattering}

\author{A. Bueno}
\email{a.bueno@ugr.es}
\author{D. Garcia-Gamez}
\email{dgarciag@ugr.es}
\author{C. Martin-Morales}
\email{cmartinmorales@ugr.es}
\affiliation{Departamento de F{\'\i}sica Te\'orica y del
  Cosmos, Universidad de Granada, E-18071 Granada, Spain}

\date{\today}

\begin{abstract} 
Two-nucleon emission is a leading uncertainty in neutrino--nucleus
interaction modeling, and no neutrino measurement constrains its
short-range-correlation (SRC) component. We propose such
a measurement: the ratio of neutral-current to charged-current
two-proton production in neutrino--argon interactions as a
function of the proton-pair relative momentum. Charged-current
interactions convert the abundant neutron--proton SRC pairs into
visible proton pairs, whereas neutral-current interactions lack an
analogous SRC contribution because proton--proton pairs are about
twenty times less abundant. The ratio is therefore predicted to
decrease at large relative momentum, where the charged-current SRC
contribution dominates, while the neutral-current sample provides a
smooth reference determined mainly by final-state interactions and
multinucleon processes. Simulations predict a suppression whose
magnitude scales with the SRC fraction. Flux normalization and
correlated detector systematics largely cancel in the ratio, making
the measurement feasible with existing liquid-argon time projection
chambers. Independently of the precise SRC contribution, this
observable provides the first direct experimental probe of the isospin
structure of weak two-nucleon emission.
\end{abstract}

\keywords{short-range correlations; neutrino--argon scattering;
neutral current; charged current; liquid-argon TPC}

\maketitle
\section{Introduction}
\label{sec:intro}

The next generation of long-baseline neutrino oscillation experiments,
DUNE~\cite{Abi:2020evt} and Hyper-Kamiokande~\cite{Abe:2018uyc}, aims
to determine the neutrino mass ordering and measure the leptonic
CP-violating phase $\delta_{\mathrm{CP}}$
with unprecedented sensitivity. Achieving these goals requires a major
reduction of the uncertainties associated with neutrino--nucleus
interaction modeling, which currently limit neutrino-energy
reconstruction and oscillation-parameter
extraction~\cite{Formaggio:2013kya, AlvarezRusoetal2018}.

At neutrino energies of order $1~\mathrm{GeV}$, several reaction
mechanisms contribute to the observed hadronic final state:
quasielastic~(QE) scattering, multinucleon excitations driven by
meson-exchange currents~(MEC)~\cite{Nieves:2011pp, Martini:2009uj},
final-state interactions~(FSI), and short-range
correlations~(SRC)~\cite{Hen:2016kwp}. Disentangling them is
challenging because they can generate similar visible hadronic
topologies.

Liquid argon time projection chamber~(LArTPC) experiments such as
ArgoNeuT~\cite{Acciarri:2014eit}, MicroBooNE~\cite{Acciarri:2016smi},
ICARUS~\cite{Acciarri:2015uup}, and
SBND~\cite{Acciarri:2015uup, Blake:2025sbnd} reconstruct
multi-nucleon final states with low thresholds and full calorimetry,
making it possible to build observables from the complete kinematics
of the hadronic system.

Short-range correlations arise from the repulsive core and tensor
component of the nucleon--nucleon ($NN$) interaction and generate
nucleon pairs with momenta well above the Fermi momentum
($k_F \approx 250~\mathrm{MeV}/c$ for $^{40}$Ar). Electron-scattering
measurements of two-nucleon knockout show that these pairs are
predominantly neutron--proton~($np$): $np$ pairs outnumber $pp$ or
$nn$ pairs in the high-momentum region by a factor of about
20~\cite{Subedi:2008zz, Duer:2018ssp, Hen:2016kwp}.

The ArgoNeuT collaboration reported back-to-back proton
configurations in charged-current interactions qualitatively
consistent with SRC-induced nucleon knockout~\cite{Acciarri:2014eit}.
Subsequent theoretical work showed that meson-exchange currents and
intranuclear rescattering, notably pion absorption on correlated
pairs, can produce topologically similar
signatures~\cite{Nieves:2011pp, Martini:2009uj, Weinstein:2016fud}.
Microscopic calculations of SRC-induced two-nucleon knockout in
neutrino scattering find the same back-to-back topology whether the
two-body current is of SRC or MEC origin~\cite{VanCuyck:2016fab},
and a NuWro analysis of the ArgoNeuT sample found that the
reconstructed back-to-back excess is kinematically expected without
additional SRC dynamics~\cite{Niewczas:2015iea}, limiting the
discriminating power of purely topological observables.

Inclusive NC/CC ratios
have constrained the axial nucleon coupling and tested lepton
universality~\cite{Formaggio:2013kya}, but they integrate over all
hadronic configurations and carry no sensitivity to the microscopic
origin of multi-proton final states. Transverse kinematic
imbalance~(TKI) observables~\cite{Lu:2015hea, Dolan:2018sbb} probe
nuclear effects differentially, but in single-proton CC topologies.
To our knowledge, no observable has combined NC/CC separation,
exclusive two-proton final states, and SRC-sensitive kinematic
variables at once.

In this paper we propose the ratio of NC to CC double-proton
production rates, $\RR$, measured differentially in $\prel$ and $\pcm$, as an
observable sensitive to SRC-enhanced charged-current dynamics. The
working hypothesis is that the NC two-proton sample is dominated by
final-state interactions and multinucleon processes and therefore
stays comparatively smooth across the ($\prel$, $\pcm$) plane. CC
interactions, by contrast, convert $np$-SRC pairs into visible
two-proton final states. A suppression of $\RR$ at large relative
momentum then arises from the SRC enhancement of the denominator, not
from a depletion of the numerator. The aim is to identify these
differential enhancements and to quantify them as the SRC fraction of
the charged-current two-proton sample (Sec.~\ref{sec:observable}).
That fraction is a property of the reconstructed sample, not of the
nuclear ground state. The pair abundances of the ground state are not
extracted directly, since final-state interactions and detection
thresholds stand between them and the observed yields. Throughout, the
expected behavior is illustrated with the NuWro event
generator~\cite{Golan:2012wx}, with the simulation setup specified in
Sec.~\ref{sec:sbnd}.

The measurement also constrains a quantity that does not depend on
which mechanism is ultimately found to produce the suppression. At the primary vertex, both
SRC breakup and multinucleon currents feed the charged-current
two-proton channel mainly through $np$ pairs, whereas the
corresponding neutral-current final state can only originate from
$pp$ configurations. Final-state interactions blur this selection.
Rescattering produces two-proton states from single-nucleon events in
both samples, and charge exchange promotes $np$ final states into the
neutral-current sample. The vertex-level isospin asymmetry therefore
reaches the measured ratio through a final-state correction that must
be taken from transport models and bounded by comparisons among them.
With that correction applied, the ratio constrains the isospin
composition of two-nucleon emission, whatever mixture of
pair-emission mechanisms produces it. This composition is not constrained by
existing inclusive data and enters the neutrino--antineutrino
cross-section asymmetries on which accelerator-based CP measurements
rest. We return to this point in Sec.~\ref{subsec:mec_separation}.

The paper is organized as follows.
Section~\ref{sec:theory} presents the theoretical framework and
defines the observable. Section~\ref{sec:observable} selects the
working projection and introduces the shape observable and the
two-component extraction of the SRC fraction.
Section~\ref{sec:sbnd} discusses experimental
considerations and presents generator-level results.
Section~\ref{sec:discussion} contains a broader discussion and
outlines future extensions. Section~\ref{sec:conclusions} summarizes
our conclusions.

\section{Theoretical Framework}
\label{sec:theory}

\subsection{Short-range correlations and their kinematics}
\label{subsec:src}

At large relative momentum and moderate center-of-mass momentum, the
two-body momentum distribution factorizes into a universal
short-distance part and the center-of-mass motion of the pair, with
weights $C_{\alpha A}$ ($\alpha = np,\,pp,\,nn$) that parametrize the
abundance of SRC pairs of each channel in a nucleus
$A$~\cite{Hen:2016kwp, Arrington:2011xs, Benharetal2008,
Wiringa:2013ala}.

The relative and center-of-mass momenta of the reconstructed proton
pair are defined from the two observed proton momenta
$\bm{p}_1,\bm{p}_2$ as
\begin{align}
\bm{\prel} &= \tfrac{1}{2}(\bm{p}_1 - \bm{p}_2)\,,
\label{eq:prel_def}\\
\bm{\pcm}  &= \bm{p}_1 + \bm{p}_2\,.
\label{eq:pcm_def}
\end{align}
With this convention $|\bm{\prel}|$ is the momentum of either nucleon
in the pair rest frame.

These reconstructed momenta are not the intrinsic momenta of the
initial pair. In the impulse approximation the probe transfers
three-momentum $\bm{q}$ to the struck nucleon, so
\begin{equation}
\bm{\prel} \simeq \bm{\krel} + \tfrac{1}{2}\bm{q}\,,
\qquad
\bm{\pcm} \simeq \bm{\kcm} + \bm{q}\,,
\label{eq:reco_shift}
\end{equation}
with $\bm{\krel},\bm{\kcm}$ the relative and center-of-mass momenta of
the initial correlated pair. We return to the consequences of this
shift in Sec.~\ref{subsec:isospin}.

Electron scattering established that SRC configurations have
$|\bm{\krel}| \gtrsim k_F \approx 250$--$300~\mathrm{MeV}/c$ and
$|\bm{\kcm}| \lesssim k_F$~\cite{Hen:2016kwp, Arrington:2011xs,
Cohen2018SRCcm}. In this region the tensor force leads to a strong
dominance of $np$ pairs. Measurements of $(e,e'pn)$ and $(e,e'pp)$
reactions found
\begin{equation}
\mathcal{R}_{np/pp}
=
\frac{C_{np,A}}{C_{pp,A}}
\approx 18\text{--}20
\label{eq:np_pp_ratio}
\end{equation}
for medium and heavy nuclei~\cite{Subedi:2008zz, Duer:2018ssp}.

In neutrino scattering, SRC configurations are not directly
observable. They manifest through the topology and kinematics of the
hadronic final state, in ways that differ between the charged- and
neutral-current channels (Sec.~\ref{subsec:isospin}).

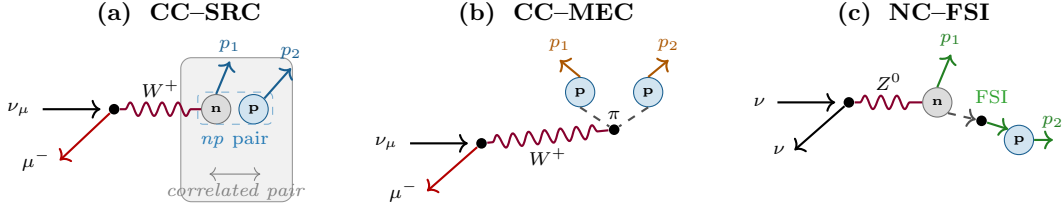
\begin{figure*}[tp]
  \centering
\begin{tikzpicture}[scale=0.9,
  proton/.style={draw=srcblue!80!black, fill=srcblue!20,
                 circle, minimum size=11pt, inner sep=0pt, font=\tiny\bfseries},
  neutron/.style={draw=darkgray!80, fill=darkgray!20,
                  circle, minimum size=11pt, inner sep=0pt, font=\tiny\bfseries},
  vertex/.style={circle, fill=black, inner sep=1.4pt},
  muon/.style={draw=red!70!black, thick, ->},
  nu/.style={draw=black, thick, ->},
  boson/.style={draw=purple!70!black, thick, decorate,
    decoration={snake, amplitude=2.5pt, segment length=6pt,
                post length=2pt, pre length=2pt}},
  nucleusbox/.style={draw=darkgray!60, rounded corners=4pt,
                     fill=darkgray!8, minimum width=42pt,
                     minimum height=54pt},
]

\begin{scope}[shift={(0,0)}]
\node[font=\small\bfseries] at (0, 1.70) {(a)\; CC--SRC};

\node[nucleusbox] (nuc1) at (0.85,0) {};
\node[neutron] (n1)  at (0.55, 0.30) {n};
\node[proton]  (p1a) at (1.10, 0.30) {p};
\node[font=\scriptsize, srcblue] at (0.82, -0.15) {\textit{np} pair};
\draw[srcblue!60, dashed, rounded corners=2pt]
  (0.28,0.08) rectangle (1.38,0.55);

\node[font=\scriptsize, anchor=east] at (-2.05, 0.30) {$\nu_\mu$};
\draw[nu] (-2.00,0.30) -- (-1.15,0.30);
\node[vertex] (v1) at (-0.95, 0.30) {};
\draw[boson] (v1) -- (n1) node[midway, above=1pt, font=\scriptsize]{$W^+$};

\draw[muon] (v1) -- (-1.75, -0.45)
  node[font=\scriptsize, left]{$\mu^-$};

\draw[->, srcblue!80!black, thick] (n1.north) -- (0.75, 1.00)
  node[font=\scriptsize, above]{$p_1$};
\draw[->, srcblue!80!black, thick] (p1a.north east) -- (1.65, 0.90)
  node[font=\scriptsize, above]{$p_2$};

\draw[<->, gray] (0.45,-0.65) -- (1.20,-0.65)
  node[midway, below, font=\scriptsize\itshape]{correlated pair};
\end{scope}

\begin{scope}[shift={(5.4,0)}]
\node[font=\small\bfseries] at (0, 1.70) {(b)\; CC--MEC};

\node[proton] (na2) at (0.50, 0.55) {p};
\node[proton] (nb2) at (1.50, 0.55) {p};
\node[vertex] (pi2) at (1.00, 0.00) {};

\draw[darkgray, dashed, thick] (na2.south) -- (pi2);
\draw[darkgray, dashed, thick] (nb2.south) -- (pi2);
\node[font=\scriptsize] at (1.00, 0.22) {$\pi$};

\node[font=\scriptsize, anchor=east] at (-2.05, -0.20) {$\nu_\mu$};
\draw[nu] (-2.00,-0.20) -- (-1.15,-0.20);
\node[vertex] (vmec) at (-0.95,-0.20) {};
\draw[boson] (vmec) -- (pi2) node[midway, below=1pt, font=\scriptsize]{$W^+$};
\draw[muon] (vmec) -- (-1.75,-0.90)
  node[font=\scriptsize, left]{$\mu^-$};

\draw[->, mecorange!80!black, thick] (na2.north) -- (0.20,1.02)
  node[font=\scriptsize, above]{$p_1$};
\draw[->, mecorange!80!black, thick] (nb2.north) -- (1.80,1.02)
  node[font=\scriptsize, above]{$p_2$};
\end{scope}

\begin{scope}[shift={(10.8,0)}]
\node[font=\small\bfseries] at (0, 1.70) {(c)\; NC--FSI};

\node[neutron] (nprim)  at (0.35, 0.40)  {n};
\node[proton]  (pfsi)   at (1.55, -0.15) {p};
\node[vertex]  (fsivtx) at (1.00, 0.13) {};

\node[font=\scriptsize, anchor=east] at (-2.05, 0.40) {$\nu$};
\draw[nu] (-2.00,0.40) -- (-1.15,0.40);
\node[vertex] (vnc) at (-0.95,0.40) {};
\draw[boson] (vnc) -- (nprim) node[midway, above=1pt, font=\scriptsize]{$Z^0$};
\draw[nu] (vnc) -- (-1.75,-0.30) node[font=\scriptsize, left]{$\nu$};

\draw[->, darkgray, thick, dashed] (nprim.south east) -- (fsivtx);
\draw[->, fsigreen!80!black, thick] (fsivtx) -- (pfsi.north west);
\node[font=\scriptsize, fsigreen] at (1.15, 0.48) {FSI};

\draw[->, fsigreen!80!black, thick] (nprim.north) -- (0.55, 1.10)
  node[font=\scriptsize, above]{$p_1$};
\draw[->, fsigreen!80!black, thick] (pfsi.east) -- (2.05, -0.15)
  node[font=\scriptsize, above=1pt]{$p_2$};
\end{scope}

\end{tikzpicture}

\caption{Mechanisms contributing to two-proton final states in
neutrino--nucleus scattering.
(a)~CC--SRC: a charged-current interaction converts the neutron member
of a correlated $np$ pair into a proton, while the correlated proton
is emitted as a recoil partner, producing two protons with large
relative momentum.
(b)~CC--MEC: multinucleon emission induced by meson-exchange currents.
(c)~NC--FSI: a neutral-current interaction followed by intranuclear
rescattering generates an additional visible proton.}
\label{fig:mechanisms}
\end{figure*}

Figure~\ref{fig:mechanisms} summarizes representative mechanisms
leading to two-proton final states. Their relative importance differs
between the CC and NC samples. This difference motivates the use of
differential kinematic observables.

\subsection{The NC/CC double-proton ratio}
\label{subsec:definition}

We define the double-differential NC/CC ratio as
\begin{equation}
\RR(\prel, \pcm) \equiv
  \frac{d^2\sigma_\mathrm{NC}(2p)/d\prel\,d\pcm}
       {d^2\sigma_\mathrm{CC}(2p)/d\prel\,d\pcm}\,,
\label{eq:ratio_def}
\end{equation}
where the numerator (denominator) refers to neutral-current
(charged-current) interactions producing exactly two reconstructed
protons with no additional pion candidates. The CC sample is
identified by the additional presence of a reconstructed muon. Both
$\prel$ and $\pcm$ are built from the two proton momenta alone, so the
observable requires no reconstruction of the incident neutrino
energy, a known source of model-dependent
bias~\cite{Lalakulich:2012cj}. This is essential for the neutral-current sample, where $E_\nu$
cannot be reconstructed, and it lets numerator
and denominator be compared bin by bin in the same reconstructed
variables.

A singly integrated ratio would average away the kinematic dependence
that distinguishes the production mechanisms. The differential
measurement in $\prel$ and $\pcm$ retains it. Being a ratio, $\RR$ also cancels the neutrino-flux
normalization exactly and correlated detector systematics largely drop
out, leaving the muon-identification efficiency as the dominant
residual. Section~\ref{subsec:syst} quantifies these cancellations.

\subsection{Physical interpretation}
\label{subsec:isospin}

\paragraph{Charged-current channel.}
The charged-current vertex $\nu_\mu\, n \to \mu^-\, p$ couples to the
neutron member of an $np$-SRC pair. When the correlated proton is
emitted above detection threshold, the final state contains two
visible protons. Because $np$ pairs dominate the SRC sector of the
nuclear wave function, CC interactions efficiently generate SRC-like
two-proton topologies. The enhancement is most visible at large
relative momentum, where the SRC component dominates the nuclear
momentum distribution.

\paragraph{Neutral-current channel.}
The neutral-current vertex conserves nucleon identity. Two visible
protons at the primary vertex require the rarer $pp$ pairs, and models
attribute the NC two-proton yield largely to final-state interactions
and multinucleon processes. These populate broad regions of phase
space. The NC sample therefore acts as a kinematic reference rather
than a clean probe of SRC abundances, and a suppression of $\RR$ in
the SRC-enhanced region reflects the growth of the CC yield where
SRC-induced proton knockout becomes important, not a depletion of NC
events.

\paragraph{Isospin expectation for the SRC region.}
The suppression of $\RR$ at large $\prel$ does not rely on the event
generator. Its leading origin is the measured isospin structure of
SRC pairs, with the observed magnitude further shaped by final-state
interactions, multinucleon currents, and detection thresholds. In the
impulse approximation a neutrino strikes the leading nucleon of an SRC
pair and the partner recoils as a spectator. The selection rules
above then admit one $2p$ configuration per $np$ pair (the struck
neutron) and two per $pp$ pair (either proton). Counting
configurations, the primary-vertex ratio (before final-state
interactions) is
\begin{equation}
\RR^{\rm SRC}
= \frac{2\,C_{pp}\,\sigma^{\nu p}_{\rm NC}}
       {C_{np}\,\sigma^{\nu n}_{\rm CC}}
= \frac{2}{C_{np}/C_{pp}}\,
  \frac{\sigma^{\nu p}_{\rm NC}}{\sigma^{\nu n}_{\rm CC}}\,,
\label{eq:isospin_floor}
\end{equation}
where $\sigma^{\nu p}_{\rm NC}$ and $\sigma^{\nu n}_{\rm CC}$ are the
elastic neutral-current and quasielastic charged-current cross
sections on a single nucleon. With the measured pair ratio
$C_{np}/C_{pp}\approx 18$--$20$~\cite{Subedi:2008zz, Duer:2018ssp} and
the measured cross-section ratio
$\sigma^{\nu p}_{\rm NC}/\sigma^{\nu n}_{\rm CC}\approx 0.15$ at BNB
energies~\cite{Ahrens:1986xe}, Eq.~(\ref{eq:isospin_floor}) gives
$\RR^{\rm SRC}\approx 0.015$--$0.017$. This lies an order of magnitude
below the non-SRC baseline ($\RR\approx0.2$) and sets a floor on the
ratio in the SRC region. The observed suppression is milder than this
floor because final-state charge exchange promotes the abundant $np$
pairs into the neutral-current $2p$ sample and, together with
multinucleon currents, partially repopulates the numerator.

Equation~(\ref{eq:isospin_floor}) provides an
analytic benchmark for the primary-vertex SRC
contribution. It is a pair-counting estimate in the impulse
approximation, neglecting interference, and evaluated before
final-state interactions and detection effects. Within these
approximations it is fixed by measured inputs (the pair ratio
$C_{np}/C_{pp}$ and the single-nucleon cross-section ratio), with no
Monte Carlo SRC prescription entering. Because NuWro does not
implement the $pp$-SRC spectral function, it cannot reproduce the
analytic benchmark of
Eq.~(\ref{eq:isospin_floor}). Its pure-SRC NC~$2p$ yield, and hence
its pure-SRC ratio, are identically zero, implying that the generator
can only \emph{underestimate} the genuine SRC contribution to the
numerator. The finite NC~$2p$ rate it does produce is FSI driven. In
a dedicated SRC sample with FSI enabled and MEC disabled, NuWro gives
$\RR\lesssim0.1$ across the accessible range, falling to
$\RR\approx0.03$ at large $\prel$. This is within a factor of two of
the analytic benchmark, consistent with the isospin counting. Since
this FSI-driven ratio already exceeds the primary-vertex benchmark of
Eq.~(\ref{eq:isospin_floor}), the NC~$2p$ yield in the SRC channel is
dominated by rescattering rather than by direct $pp$-SRC breakup, and
the omission of the $pp$-SRC spectral function biases the numerator by
less than the FSI contribution itself. The default
sample, in which meson-exchange currents populate both channels, sits
higher (Sec.~\ref{subsec:stats}). Removing this common contribution
exposes the SRC-driven suppression, which settles close to the
benchmark of Eq.~(\ref{eq:isospin_floor}).

The suppression therefore traces to measured pair abundances, not to
the generator's SRC prescription. The SRC-fraction parameter controls the
$np$-SRC breakup that enhances the CC denominator, while the NC
numerator, lacking the $pp$-SRC channel, cannot generate a comparable
primary-vertex contribution. The observed depth of the suppression,
however, remains shaped by final-state charge exchange and other
transport effects. The residual gap between the analytic benchmark
and the NuWro prediction is therefore expected to be dominated by the
modeling of charge exchange. This localizes the remaining model
dependence in a single transport ingredient, whose treatment across
independent calculations is discussed in Sec.~\ref{subsec:fsi}.

In electron scattering, SRC pairs are tagged by their low $\kcm$. In a
neutrino interaction, Eq.~(\ref{eq:reco_shift}) shifts both
reconstructed momenta by the transfer $\bm{q}$, whose magnitude is
typically several hundred $\mathrm{MeV}/c$, and the two shifts are not
equivalent. The center-of-mass momentum $\pcm \simeq \kcm + \bm{q}$
is dominated by $\bm{q}$, because $\kcm \lesssim k_F$ for every
mechanism (SRC pairs and mean-field pairs alike). The reconstructed
$\pcm$ distribution is therefore broad and similar for all two-proton
production processes, and it carries little SRC-discriminating power.
The relative momentum $\prel \simeq \krel + \tfrac{1}{2}\bm{q}$, by
contrast, retains the large $\krel$ that distinguishes SRC pairs from
all other configurations. The SRC signature is thus washed out in
$\pcm$ but preserved as a high-$\prel$ tail, and the SRC-driven
variation of $\RR$ appears in $\prel$ rather than in $\pcm$. The
generator-level distributions of Sec.~\ref{subsec:stats}
(Fig.~\ref{fig:kinematics}) confirm both statements.

\subsection{Meson-exchange currents}
\label{subsec:mec}

Meson-exchange currents generate multinucleon emission without a
pre-existing SRC configuration~\cite{Nieves:2011pp, Martini:2009uj}.
Within a single sample, MEC can populate the large-$\prel$ region and
mimic the SRC signature. In 2p2h emission
both outgoing nucleons share the transferred energy--momentum, so the
relative momentum of the final pair is set by the energy sharing
rather than by any initial-state correlation. Exclusive 2p2h
calculations show that the emitted pairs extend to large relative
momenta and back-to-back configurations~\cite{RuizSimo:2016mec,
Sobczyk:2020dhh, Kasturi:2026exc}. A cut on $\prel$ alone therefore
does not isolate SRC breakup.

The ratio structure of the observable protects against this, but the
argument must be made with care, because MEC shares with SRC the
dominance of $np$ pairs in charged-current
interactions~\cite{RuizSimo:2016pairs} and is therefore subject to the same channel
selection (Sec.~\ref{subsec:isospin}). One could then object that MEC feeds
the CC tail but not the NC tail, exactly as SRC does, and that the
ratio cannot separate the two mechanisms. This objection does not
distinguish between the \emph{rate} of a mechanism and the
\emph{shape} of its $\prel$ distribution. The energy sharing that
sets the pair's relative momentum is the same whether the initial
pair was $np$ or $pp$. The $np$
dominance of MEC therefore suppresses the NC-MEC rate by a factor that
is approximately independent of $\prel$: it removes NC-MEC events
uniformly, not preferentially from the tail. For SRC the situation is
different. The high-$\prel$ events originate in the high-momentum
component of the initial-state pair distribution, which only
correlated pairs possess, and the same tensor force that generates
this component makes those pairs predominantly $np$. For SRC, and
only for SRC, the isospin asymmetry and the concentration in the tail
have a common origin.

To make this quantitative, we decompose the ratio over production
mechanisms $m \in \{\mathrm{SRC, MEC, FSI}\}$,
\begin{equation}
\RR(\prel) \;=\;
\frac{\sum_m r_m(\prel)\, w_m(\prel)}{\sum_m w_m(\prel)}\,,
\qquad
r_m \equiv \frac{\mathrm{NC}_m}{\mathrm{CC}_m}\,,
\label{eq:mech_decomposition}
\end{equation}
where $w_m(\prel)$ is the CC yield of mechanism $m$ and $r_m$ its
channel asymmetry. The sum is incoherent: cross sections are added,
not amplitudes. Event generators implement these mechanisms in the
same way, as separate processes without interference terms. In a
microscopic calculation,
correlation and exchange currents interfere and their separation is
scheme dependent, so the labels are operational: SRC denotes the
component whose CC weight is concentrated in the tail and whose
channel asymmetry is strongly suppressed. The arguments below use
these two properties, not the label. Equation~(\ref{eq:mech_decomposition})
admits a slope of $\RR$ through two routes only: a $\prel$ dependence
of one of the asymmetries $r_m$, or a change with $\prel$ of the
mechanism composition among components with different $r_m$. The SRC
signal proceeds through the second route: $w_{\mathrm{SRC}}$ grows in
the tail, where the initial-state high-momentum component resides,
while $r_{\mathrm{SRC}}$ lies far below the background asymmetries.

For MEC to produce a comparable fall, either $r_{\mathrm{MEC}}$ must
decrease substantially with $\prel$, or the MEC contribution must
combine a strongly suppressed $r_{\mathrm{MEC}}$ with a CC-MEC
$\prel$ shape harder than that of the other backgrounds. Both are
specific, testable model statements, and neither is expected, for the
following reasons. In microscopic 2p2h calculations the $np$-to-$pp$
emission ratio is driven mainly by the $\Delta$ current and depends on
the energy and momentum transfer $(\omega,q)$~\cite{RuizSimo:2016pairs}, so some
$\prel$ dependence of $r_{\mathrm{MEC}}$ cannot be excluded a priori.
This dependence is however diluted in the projection onto pair
kinematics: at fixed $(\omega,q)$ the energy sharing spreads the
emitted pair over a broad range of $\prel$, so
$r_{\mathrm{MEC}}(\prel)$ is a smeared average of the underlying
$(\omega,q)$ variation and is flatter than it. Recent semi-exclusive
calculations give the size of the residual effect: the relative yield
of $pp$ and $np$ pairs changes by up to ${\sim}20\%$ between kinematic
regions and between treatments of the nuclear ground
state~\cite{Rocco:2026two}. This variation is five times smaller than
the factor-of-two (100\%) modulation that would be needed to reproduce
the SRC suppression even if MEC dominated the background. No isospin-resolved exclusive 2p2h
prediction exists for the neutral-current channel, whose multinucleon
component event generators still treat empirically. The NC side of
$r_{\mathrm{MEC}}$ is an extrapolation in every current model. This
limitation cuts both ways: it motivates the in-situ sideband strategy
of Sec.~\ref{subsec:mec_separation}, and it makes the measurement
itself the first constraint in this sector.

The same considerations
apply to the FSI component, whose channel asymmetry could in principle
vary with the pair kinematics through the charge-exchange probability.
This variation is probed by the same SRC-disabled configuration and by
the comparisons among transport models discussed in
Sec.~\ref{subsec:syst}. The generator study of Sec.~\ref{sec:sbnd}
supports the flat expectation (open circles in Fig.~\ref{fig:ratio}):
with the SRC model disabled but MEC and FSI fully active, both samples
populate the entire $\prel$ range and their ratio shows no $\prel$
dependence. This flatness is a model expectation supported by the
simulation, not a theorem. It could even be partly built in: NuWro
generates 2p2h final states essentially by phase space, and its
intranuclear cascade carries no initial-state correlations, so a
smooth tail is a natural output of the sampling. For this reason the
proposal treats the flatness as a quantity to be measured rather than
assumed, in the sidebands of Sec.~\ref{subsec:mec_separation}, and to
be checked against the exclusive 2p2h
calculations~\cite{Sobczyk:2020dhh, Rocco:2026two, Kasturi:2026exc}.
Its validation is the main remaining model dependence of the proposal
(Sec.~\ref{subsec:syst}).

\subsection{Final-state interactions}
\label{subsec:fsi}

For two-proton topologies, final-state interactions generate
additional visible protons, distort the pair kinematics, and migrate
events across the $(\prel,\pcm)$ plane. Reconstructed SRC-like
configurations therefore do not correspond uniquely to primary SRC
breakup at the vertex.

For the proposed observable, FSI play a dual role. They contribute
significantly to the NC double-proton sample, defining the reference
behavior of the numerator, and they smear the kinematics of CC
SRC-like events. A realistic interpretation of $\RR(\prel,\pcm)$
therefore requires detailed transport modeling, provided here by
NuWro. Since generators implement SRC, MEC, and FSI
dynamics in distinct ways~\cite{Alvarez-Ruso:2021oui, Golan:2012wx,
Buss:2011mx}, a comparison across independent generators is a
necessary validation step. Its scope, however, is bounded. The
definition of the observable, the in-situ measurement of the
background reference, and the isospin floor on the SRC channel
asymmetry are independent of the transport model. What the comparison
must constrain is one quantity: the charge-exchange repopulation of
the NC numerator, which sets the upper edge of the SRC reference
value used in the extraction of Sec.~\ref{sec:observable} and shifts
the extracted fraction only weakly. A meaningful comparison also requires generator configurations
in which the SRC component can be switched off and rescaled
consistently, which not all generators expose. We therefore leave it
to a dedicated study, noting that it affects the interpretation of
the measured depth, not the design or the feasibility of the
measurement.

\section{Working Observable and SRC-Fraction Extraction}
\label{sec:observable}

The discriminating information is not spread evenly over the
$(\prel,\pcm)$ plane.
As argued in Sec.~\ref{subsec:isospin}, the momentum transfer dominates
the reconstructed $\pcm$, so the SRC-driven variation of $\RR$ appears
in $\prel$ and not in $\pcm$. We therefore adopt the projection
$\RR(\prel)$, evaluated over a common $\pcm$ acceptance, as the
working observable. The $\pcm$ direction is retained as a consistency
check.

Because MEC and FSI act mainly on the overall level of the ratio while
the SRC mechanism generates its fall with $\prel$
(Sec.~\ref{subsec:mec}), the most model-robust quantity is the
\emph{shape} of $\RR(\prel)$. We therefore also define the normalized
ratio
\begin{equation}
\DR(\prel) \;\equiv\;
\frac{\RR(\prel)}{\big\langle \RR \big\rangle_{\mathrm{norm}}}\,,
\label{eq:double_ratio}
\end{equation}
where $\langle \RR \rangle_{\mathrm{norm}}$ is the ratio averaged over
a normalization window at moderate relative momentum
($0.3 \lesssim \prel \lesssim 0.5~\mathrm{GeV}/c$), where the SRC
contribution is small. In $\DR$ every $\prel$-independent factor
cancels: residual flux-shape effects, the muon-identification
efficiency (the dominant non-canceling detector systematic of $\RR$,
cf.~Sec.~\ref{subsec:syst}), and the overall MEC and FSI normalization
of both samples. The choice of window is model-informed and must be
varied as part of the systematic assessment. A residual SRC
contribution inside the window rescales $\DR$ by a constant and does
not affect its fall with $\prel$, which carries the signal. The
construction thus reduces the measurement to a shape comparison in
which most level uncertainties drop out.

The same two-component logic gives the measurement a quantitative,
weakly model-dependent interpretation. If the CC sample at
given $\prel$ consists of an SRC component and a background of all
other mechanisms, with channel asymmetries $r_{\mathrm{SRC}}$ and
$r_{\mathrm{bg}}$ [Eq.~(\ref{eq:mech_decomposition})], then
\begin{equation}
\RR(\prel) \;=\;
\frac{r_{\mathrm{bg}} + r_{\mathrm{SRC}}\, S(\prel)}{1 + S(\prel)}\,,
\qquad
S \equiv \frac{\mathrm{CC}_{\mathrm{SRC}}}{\mathrm{CC}_{\mathrm{bg}}}\,,
\label{eq:two_component}
\end{equation}
which inverts to
\begin{equation}
f_{\mathrm{SRC}}(\prel) \;=\; \frac{S}{1+S}
\;=\;
\frac{r_{\mathrm{bg}} - \RR(\prel)}{r_{\mathrm{bg}} - r_{\mathrm{SRC}}}\,.
\label{eq:src_fraction}
\end{equation}
The SRC fraction of the CC two-proton sample is the fractional
distance of the measured ratio from the background level toward the
SRC level. The background asymmetry $r_{\mathrm{bg}}$ is measured in
situ, from the same data, in the low-$\prel$ region where SRC pairs
are kinematically scarce. Any residual SRC contamination of that
region lowers the measured reference level and biases $f_{\mathrm{SRC}}$
downward, so the extraction is conservative, and the bias can be
removed iteratively using the extracted fraction itself. The SRC
asymmetry $r_{\mathrm{SRC}}$ is bounded from below by the isospin
floor of Eq.~(\ref{eq:isospin_floor}), which is fixed by measured
inputs within the pair-counting approximation of
Sec.~\ref{subsec:isospin}, and from above by its value after FSI repopulation, which
does require transport modeling. The band refers to the SRC-dominated
tail. At low $\prel$, where SRC events arrive only after substantial
reprocessing, the effective asymmetry rises above it and the
extraction becomes a conservative lower bound, as the closure test of
Sec.~\ref{subsec:stats} quantifies. This residual model dependence is
weak because $r_{\mathrm{SRC}} \ll r_{\mathrm{bg}}$: since
$\partial f_{\mathrm{SRC}}/\partial r_{\mathrm{SRC}} =
f_{\mathrm{SRC}}/(r_{\mathrm{bg}} - r_{\mathrm{SRC}})$, the full width
of the band shifts $f_{\mathrm{SRC}}$ by less than $0.05$ even at
$f_{\mathrm{SRC}} = 0.6$. The construction does not
require the background asymmetry to be strictly flat. Should
$r_{\mathrm{bg}}$ carry a residual $\prel$ dependence
(Sec.~\ref{subsec:mec}), it is measured as a function of $\prel$ in
the MEC-enriched sidebands of Sec.~\ref{subsec:mec_separation} and
Eq.~(\ref{eq:src_fraction}) is applied pointwise. The sensitivity to
an imperfect reference level is in any case small. Since $\partial f_{\mathrm{SRC}}/\partial
r_{\mathrm{bg}} = (\RR - r_{\mathrm{SRC}})/(r_{\mathrm{bg}} -
r_{\mathrm{SRC}})^2$, an uncorrected background tilt of $0.01$ shifts
$f_{\mathrm{SRC}}$ by only $0.02$--$0.04$. The measurement thus
returns the SRC fraction of the charged-current two-proton sample,
defined within the two-component decomposition, as
a function of $\prel$, not only the presence of a suppression, with
the two-component assumption itself testable through the kinematic
sidebands of Sec.~\ref{subsec:mec_separation}.

\section{Experimental Considerations and Generator-Level Results}
\label{sec:sbnd}

\subsection{Event selection}
\label{subsec:selection}

A LArTPC offers a low proton kinetic-energy threshold
($T_p \gtrsim 40~\mathrm{MeV}$, $p_p \gtrsim 275~\mathrm{MeV}/c$),
$dE/dx$-based particle identification, calorimetric proton energy
reconstruction at the ${\lesssim}5\%$ level for stopping tracks, and
identification of $\pi^0$ and charged-pion topologies for a pion veto.
The low proton threshold matters because both protons of the pair must
be reconstructed. The spectator proton of an SRC pair carries a
momentum of order $\krel$, so requiring it above threshold
($p_p \gtrsim 275~\mathrm{MeV}/c \approx k_F$) already selects the
high relative momenta characteristic of SRC. The threshold therefore
shapes the observable toward the SRC-sensitive region rather than
against it. We define two event samples, each requiring a
reconstructed primary vertex inside the fiducial volume.

\textbf{CC 2p sample}: one reconstructed muon candidate (momentum
$p_\mu > 100~\mathrm{MeV}/c$), exactly two reconstructed proton
candidates ($T_p > 40~\mathrm{MeV}$ each), no identified pion
candidates above $50~\mathrm{MeV}/c$.

\textbf{NC 2p sample}: no reconstructed muon candidate, exactly two
reconstructed proton candidates ($T_p > 40~\mathrm{MeV}$ each),
no identified pion candidates above $50~\mathrm{MeV}/c$.

The main background to the CC 2p sample comes from charged-current
pion-production events in which the pion is absorbed in the argon
nucleus. Such CC$\pi$ events give a pion-less two-proton topology
that mimics the signal~\cite{Weinstein:2016fud}. In $^{40}$Ar, intranuclear pion-absorption
probabilities for low-energy pions reach 20--30\%, depending on the
pion energy and the FSI model. These events contaminate the CC
denominator and dilute the genuine SRC-enhanced fraction, a leading
systematic. Similar absorbed-pion processes feed the NC selection and
must be modeled consistently in both channels. Pion sidebands and
comparisons among FSI models will be needed to bound the resulting
model dependence.

The NC sample picks up further backgrounds from CC events whose muon
is unreconstructed or fails identification, and from NC and CC pion
production with unreconstructed, subthreshold, or absorbed pions.

The muon leakage admits a parametric bound. Let $\epsilon$ be the
probability that a CC~$2p$ event loses its muon to reconstruction or
identification failure, taken independent of the pair kinematics as
in Sec.~\ref{subsec:syst}. The leaked events follow the $\prel$
distribution of the CC sample itself, so the measured ratio becomes
$(\RR+\epsilon)/(1-\epsilon)$: the contamination adds an offset
$\epsilon\,(1+\RR)$, nearly constant in $\prel$, and cannot imitate or
erase the SRC-driven fall. Each percent of misidentification raises
the ratio by ${\approx}0.01$. The extraction of
Sec.~\ref{sec:observable} absorbs most of even this offset, because
the background reference $r_{\mathrm{bg}}$ is measured from the same
contaminated data and shifts by nearly the same amount. The numerator
of Eq.~(\ref{eq:src_fraction}) is then preserved up to a correction of
order $\epsilon$, the bias acts through the denominator, and each
percent of misidentification lowers $f_{\mathrm{SRC}}$ by only
${\approx}0.03$ at $f_{\mathrm{SRC}}=0.6$, comparable to the
$r_{\mathrm{SRC}}$ band and in the conservative direction. The
reverse leakage, an NC event acquiring a fake muon candidate, is
suppressed by both the NC-to-CC ratio and the proton-to-muon
misidentification rate and enters the denominator at the per-mille
level.

The NC sample is also flavor blind: the small $\nu_e$ and wrong-sign components of
the beam contribute to the numerator at the per-cent level and are
folded into the flux-shape uncertainty. Neutral-pion backgrounds come mainly from
$\pi^0 \rightarrow \gamma\gamma$ conversions, which liquid-argon
calorimetry and topology reject well. Cosmic rays and interactions
outside the fiducial volume are removed by timing, containment, and
topology cuts, and can be constrained with sidebands.

\subsection{Expected behavior from NuWro simulations}
\label{subsec:stats}

We generate neutrino--argon interactions with the NuWro event
generator~\cite{Golan:2012wx} and apply the CC~$2p$ and NC~$2p$
selections of Sec.~\ref{subsec:selection} at the level of true
final-state protons above threshold. Multinucleon emission is modeled
with the transverse-enhancement (TEM) parametrization of two-body
currents.
The magnitude of these effects is model dependent. The purpose of the
simulation is to establish the sign of the effect, its localization in the
$(\prel,\pcm)$ plane, and its scaling with the SRC fraction
implemented in the generator, all of which a measurement can test.

\begin{figure}[tbp]
\centering
\includegraphics[width=\columnwidth]{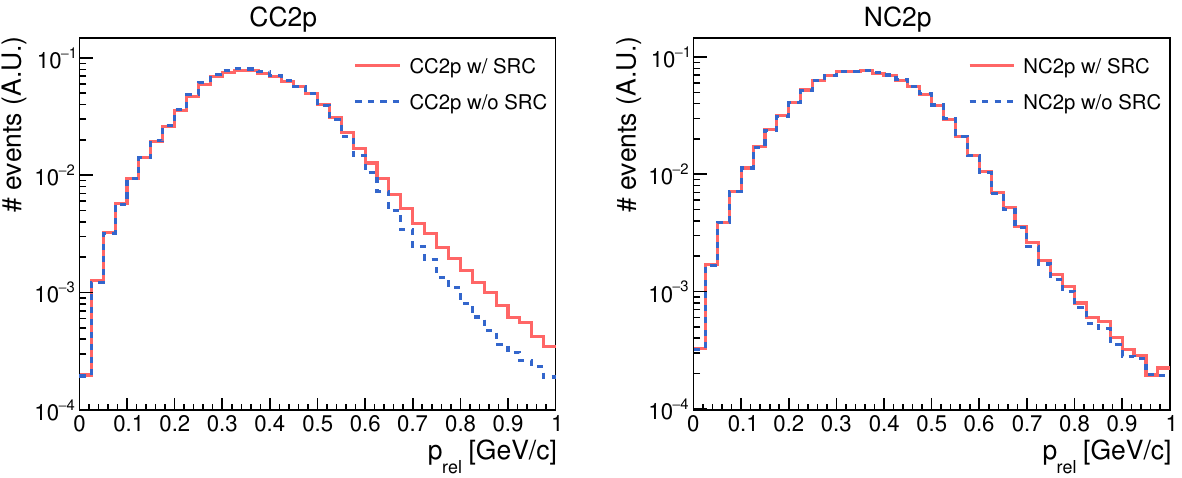}\\
\includegraphics[width=\columnwidth]{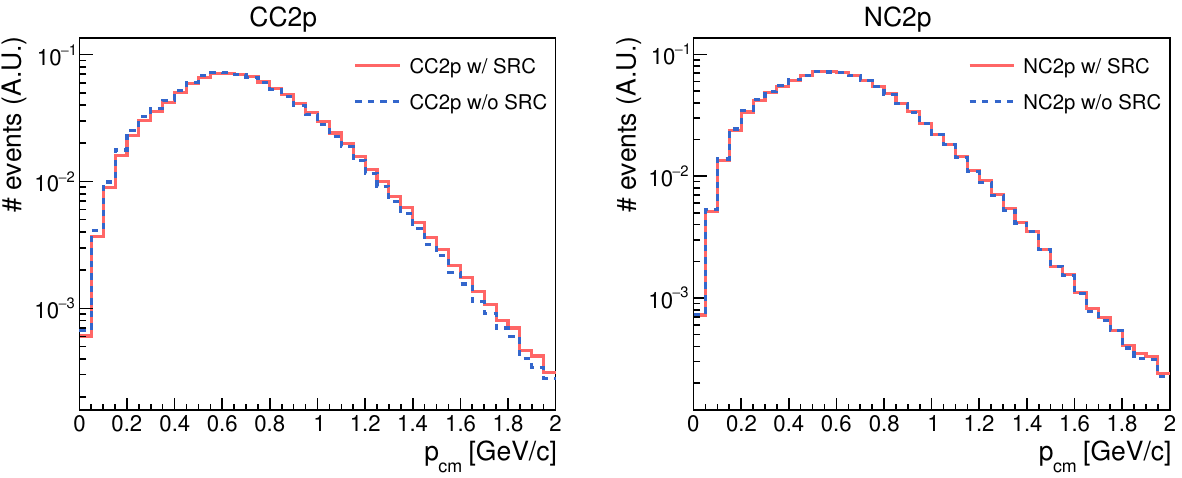}
\caption{Relative-momentum (top) and center-of-mass-momentum (bottom)
distributions of the reconstructed proton pair for the CC~$2p$ and
NC~$2p$ samples, normalized to unit area, from NuWro with the SRC
model enabled (solid) and disabled (dashed). The SRC setting affects
only the CC sample, and only as a harder tail at
$\prel \gtrsim 0.5~\mathrm{GeV}/c$. The $\pcm$ distributions are
broad and closely similar for all configurations.}
\label{fig:kinematics}
\end{figure}

Figure~\ref{fig:kinematics} (top) compares the CC~$2p$ and NC~$2p$
relative-momentum distributions, with the SRC model enabled and
disabled. Below $\prel \approx 0.5~\mathrm{GeV}/c$, where MEC- and
FSI-induced production dominates both samples, the four distributions
are nearly identical. Above this value the CC distribution with SRC
enabled develops a markedly harder tail, fed by the conversion of
$np$-SRC pairs into visible two-proton final states. Disabling the
SRC model removes the tail and the CC distribution collapses onto the
NC ones, while the NC sample is essentially insensitive to the SRC
setting. The tail is therefore charged-current specific and SRC
driven, at the level of the individual distributions and before any
ratio is formed. Because this excess appears in the
\emph{denominator} of $\RR$, it drives the ratio downward at large
$\prel$.

The corresponding $\pcm$ distributions
[Fig.~\ref{fig:kinematics} (bottom)] show the opposite behavior. All
four distributions are broad, peak in the same region, and are
essentially unchanged by the SRC setting in either sample. This is
the generator-level confirmation of the kinematic argument of
Sec.~\ref{subsec:isospin}: the reconstructed $\pcm$ is dominated by
the momentum transfer, common to all production mechanisms, so the
low-$\pcm$ SRC tag of electron scattering does not survive and the
discriminating information resides entirely in $\prel$.

\begin{figure}[tbp]
\centering
\includegraphics[width=\columnwidth]{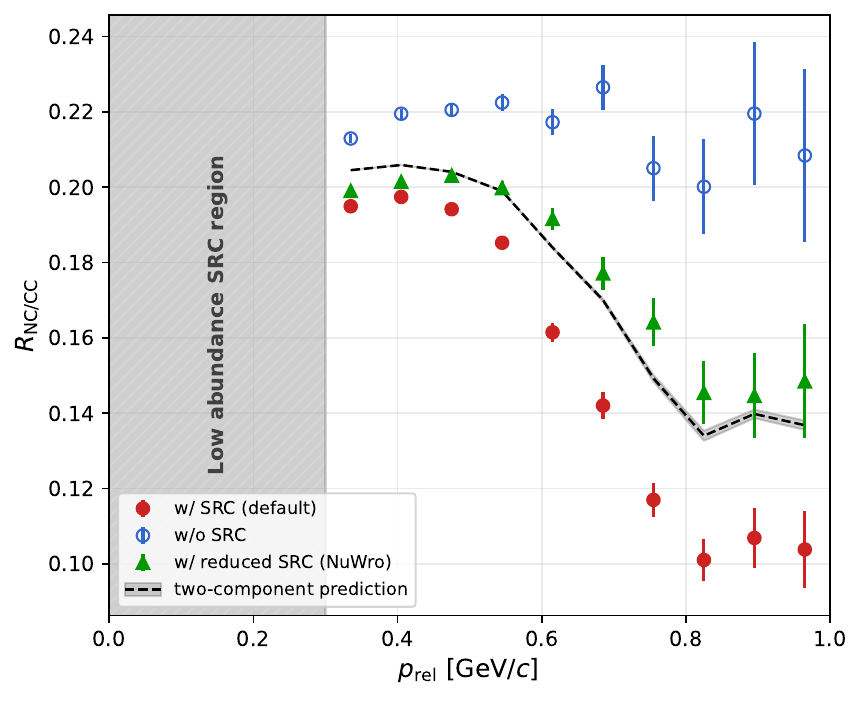}
\caption{$\RR$ as a function of $\prel$ for three NuWro
configurations: default SRC fraction (filled circles, $20\%$ of pairs
treated as short-range correlated), reduced SRC fraction (triangles,
$10\%$), and SRC model disabled (open circles, the pure MEC-plus-FSI
reference). Dashed: two-component prediction for the reduced-SRC
configuration, derived pointwise from the other two curves through
Eq.~(\ref{eq:two_component}) with no fitted parameters. The shaded
band marks the kinematic domain of low SRC-pair abundance. Error bars
are statistical (simulation).}
\label{fig:ratio}
\end{figure}

Figure~\ref{fig:ratio} shows $\RR$ as a function of $\prel$ for three
NuWro configurations that differ only in their SRC fraction. The
default setting treats $20\%$ of nucleon pairs as short-range
correlated, the reduced setting $10\%$, and the third disables the SRC
model. With the default fraction, $\RR$ falls from $\approx 0.20$
near $\prel = 0.3~\mathrm{GeV}/c$ to $\approx 0.10$ at
$\prel = 1~\mathrm{GeV}/c$. With the SRC model disabled, it stays
nearly flat
at $\approx 0.22$ across the range, and the reduced setting lies
between the two. The depth of the suppression thus scales with the
generator SRC fraction. With the SRC model disabled, MEC and FSI
still populate the full $\prel$ range (events at
$\prel \approx 1~\mathrm{GeV}/c$ are present in both samples), yet
the ratio (open circles in Fig.~\ref{fig:ratio}) exhibits nearly no $\prel$
dependence. The
presence of tail events is not by itself an SRC signal. In this
simulation, MEC and FSI do not generate a charged-current--specific
excess in the tail. They fix the \emph{level} of $\RR$, while the
\emph{shape}, its fall with $\prel$, tracks the SRC mechanism
(cf.\ Sec.~\ref{subsec:mec}). Expressed through the shape observable
of Eq.~(\ref{eq:double_ratio}), the same information reads:
normalizing each configuration to its average over the
$0.3$--$0.5~\mathrm{GeV}/c$ window leaves the SRC-disabled ratio at
unity across the full range, while the default configuration falls to
$\DR \approx 0.5$ at $\prel = 1~\mathrm{GeV}/c$. The suppression is
thus a factor-of-two shape distortion, the form in which the
measurement is least exposed to the normalization systematics of
Sec.~\ref{subsec:syst}. The shaded band at
$\prel < 0.3~\mathrm{GeV}/c$ marks the region where SRC pairs are
kinematically scarce and no suppression is expected.

The two-component form of Eq.~(\ref{eq:two_component}) describes
these results with no further input from the generator. Reading
$r_{\mathrm{bg}} \approx 0.22$ from the SRC-disabled curve and taking
$r_{\mathrm{SRC}} \approx 0.015$--$0.03$ from the isospin floor and
its value after FSI repopulation, the default-SRC point $\RR \approx 0.10$ at
$\prel = 1~\mathrm{GeV}/c$ corresponds through
Eq.~(\ref{eq:src_fraction}) to $f_{\mathrm{SRC}} \approx 0.6$. In
this description, SRC breakup accounts for roughly $60\%$ of the
CC~$2p$ sample at the highest $\prel$. The form also relates the curves to
one another. Halving the generator SRC fraction halves $S(\prel)$, so
the reduced-SRC curve is fixed pointwise by the other two,
$\RR^{\mathrm{red}} = [\,r_{\mathrm{bg}} + r_{\mathrm{SRC}}
S/2\,]/[\,1 + S/2\,]$ with $S(\prel)$ extracted from the default
curve. This derived curve, shown dashed in Fig.~\ref{fig:ratio},
involves no additional simulation and no fitted parameters, and
varying $r_{\mathrm{SRC}}$ across the band of
Eq.~(\ref{eq:isospin_floor}) moves it by less than a few per cent. It
reproduces the reduced-SRC points within the simulation statistics.
That the generator curves are related by a two-parameter formula
indicates that the simulation enters the predicted ratio mainly
through those two numbers, which in a measurement are determined by
data and isospin counting up to a residual transport correction
(Sec.~\ref{sec:observable}). The generator also shapes the samples
themselves, through final-state interactions, pion absorption, and
bin migration. Those effects enter both channels and are addressed in
Secs.~\ref{subsec:selection} and~\ref{subsec:syst}. This is
the intended role of the generator in this proposal.

\begin{figure}[tbp]
\centering
\includegraphics[width=\columnwidth]{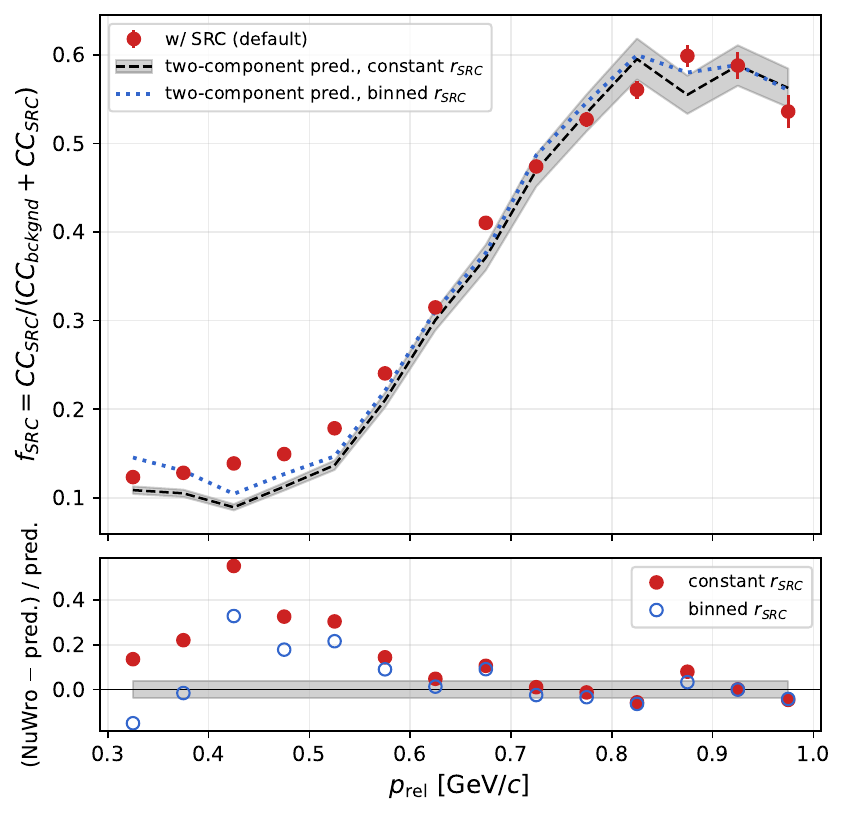}
\caption{Closure test of the two-component extraction, NuWro default
configuration. Points: true SRC fraction of the CC~$2p$ sample from
the event record. Black dashed: fraction reconstructed through
Eq.~(\ref{eq:src_fraction}), with $r_{\mathrm{bg}}$ from the
SRC-disabled configuration and constant
$r_{\mathrm{SRC}} = 0.015$--$0.03$ (band). Blue dotted: the same
with the $\prel$-binned $r_{\mathrm{SRC}}$ of the dedicated
SRC sample. Bottom: relative differences.
}
\label{fig:fsrc_closure}
\end{figure}

Figure~\ref{fig:fsrc_closure} shows a closure test of the two-component
extraction. The \emph{true} SRC fraction of the CC~$2p$ sample, known from the
generator event record, is compared with the fraction
\emph{reconstructed} through Eq.~(\ref{eq:src_fraction}) using only
the quantities available to an experiment: the measured ratio
$\RR(\prel)$ and the two reference values $r_{\mathrm{bg}}$ and
$r_{\mathrm{SRC}}$. For $\prel \gtrsim 0.6~\mathrm{GeV}/c$ the two
determinations agree within the simulation statistics and the
$r_{\mathrm{SRC}}$ band. This validates the internal consistency of
the extraction (the map from the measured ratio to the SRC fraction)
within a realistic transport environment containing MEC and FSI, in
the region where the observable is designed to operate. It does not,
and is not meant to, validate the physical correctness of the
generator's SRC or FSI modeling, whose role remains illustrative.

Below $\prel \approx 0.6~\mathrm{GeV}/c$ the true fraction exceeds
this reconstruction, by up to $\Delta f_{\mathrm{SRC}} \approx 0.05$
in absolute terms [Fig.~\ref{fig:fsrc_closure} (bottom), filled
circles]. The deviation is large in relative terms only because the
fraction itself is small there. The simulation traces it to two
$\prel$ dependences of the effective asymmetries. The first is the
SRC asymmetry. The SRC events that populate the low-$\prel$ region
are those most reprocessed by final-state interactions, and charge
exchange raises their channel asymmetry above the primary-vertex
band, to ${\approx}0.1$ in the dedicated SRC sample. Reconstructing
with this $\prel$-binned asymmetry (blue dotted in
Fig.~\ref{fig:fsrc_closure}) removes roughly half of the deviation
[open circles]. The remainder measures the background reference.
Within the SRC-enabled sample the background asymmetry acquires a
residual $\prel$ dependence that the flat SRC-disabled ratio does not
capture. Both dependences are anticipated by the construction of
Sec.~\ref{sec:observable}. A background tilt is measured pointwise
in the MEC-enriched sidebands, and FSI reprocessing can only raise
the effective SRC asymmetry above the band. In the simulation the
background tilt acts in the same direction over most of the range, so
the constant-band extraction is biased downward and the low-$\prel$
result is a conservative lower bound. The closure test thus maps the
domain of validity of the two-component description with fixed
reference values. It is quantitative in the SRC-dominated tail and
conservative below it, where the region serves in the measurement as
the background reference rather than as signal.

We have also examined $\RR$ over the full $(\prel,\pcm)$ plane in the
simulation. The SRC-driven suppression develops at large $\prel$
across the whole populated $\pcm$ range rather than in a low-$\pcm$
corner, and the two-dimensional map contains no structure beyond the
projection already shown, so we do not display it. The experimental
analysis should nevertheless perform the measurement in the full
plane, where the absence of $\pcm$ structure is itself a validation test
of the kinematic argument of Sec.~\ref{subsec:isospin}.

The error bars shown are the statistical uncertainties of the
simulated samples. Detector resolution is not folded in, and its
effect on the shape is small. The calorimetric energy scale of
Sec.~\ref{subsec:selection} corresponds to a per-proton momentum
resolution of $2$--$3\%$, and degree-level tracking angles contribute
comparably, giving $\sigma(\prel)\approx0.2$--$0.3~\mathrm{GeV}/c$ for
the pair. This is a fraction of the bin width and far below the
${\sim}0.5~\mathrm{GeV}/c$ scale over which the suppression develops,
so bin migration is a nearest-neighbor effect that dilutes rather
than reshapes the fall. Being common to both samples, the smearing
belongs to the correlated reconstruction systematics of
Sec.~\ref{subsec:syst}. Taking SBND as a concrete example of the
achievable precision, an exposure of $6.6\times10^{20}$ protons on
target corresponds to about $4\times10^{6}$ inclusive $\nu_\mu$
charged-current interactions~\cite{Blake:2025sbnd}. The exclusive
CC~$2p$ topology retains a few percent of this sample, and the
SRC-enhanced tail ($\prel \gtrsim 0.5~\mathrm{GeV}/c$) a further
${\sim}10\%$, leaving of order $10^{4}$ CC and $10^{3}$ NC events
integrated over the SRC-enhanced region. The corresponding
statistical uncertainty on $\RR$ there is at the few-percent level,
small compared with the factor-of-two suppression predicted in
Fig.~\ref{fig:ratio}. These fractions are illustrative
generator-level estimates rather than a full efficiency-folded
projection. Even so, they indicate that the measurement is not
statistics limited at existing LArTPC exposures and that its precision
will be set by systematic and modeling uncertainties.

\subsection{Systematic Uncertainties}
\label{subsec:syst}
\label{sec:systematics}

Table~\ref{tab:systematics} summarizes the main sources of systematic
uncertainty and their cancellation behavior in the ratio.

Neutrino-flux normalization cancels exactly at leading order, since
both cross sections scale linearly with the incoming flux. Some
sensitivity to the flux shape remains, from the different energy
dependences of the NC and CC cross sections and from the CC kinematic
threshold\footnote{Since the BNB flux extends below the effective CC
threshold, the NC sample integrates over a slightly broader
neutrino-energy range. This induces a mild kinematic asymmetry
between numerator and denominator even before nuclear effects are
included.}. Both effects vary slowly at energies of order 1 GeV and
cancel only partially. Detector uncertainties on proton
reconstruction (efficiency, calorimetric energy scale, threshold
modeling, and fiducial-volume corrections) affect numerator and
denominator through identical algorithms and largely cancel. The
dominant non-canceling detector systematic is muon identification
efficiency, which is unique to the CC sample and must be controlled
through dedicated data-driven methods. To the extent that this
efficiency does not depend on the proton-pair kinematics, it acts as a
pure normalization of the CC sample and cancels in the shape
observable $\DR(\prel)$ of Eq.~(\ref{eq:double_ratio}).

The largest modeling uncertainties arise from FSI and MEC. In NuWro
the SRC, MEC, and FSI contributions overlap in the $(\prel,\pcm)$
plane, so they cannot be separated by a kinematic selection within the
sample. Their uncertainties must be controlled externally. MEC
uncertainties benefit from independent measurements of CC~$0\pi$ cross
sections at MicroBooNE and T2K~\cite{Abratenko:2022nmc}, which reduce
the allowed variation of the MEC contribution relative to untuned
generator predictions. Although these measurements are inclusive in
proton multiplicity, they constrain the overall normalization and
kinematic behavior of the multinucleon contribution. For the ratio,
moreover, the relevant MEC uncertainty is not its normalization (which
affects both channels and cancels in $\DR$) but a possible
channel-asymmetric $\prel$ dependence of the emitted-pair isospin
composition (Sec.~\ref{subsec:mec}). Generator-internal exchanges of
the two-body-current model do not test this quantity, because the
alternative implementations act on the charged-current channel only
and an exchange therefore does not yield a consistent alternative
prediction for the channel asymmetry. Its approximate flatness can
instead be tested against the exclusive 2p2h calculations and measured
directly in the MEC-enriched sidebands
(Sec.~\ref{subsec:mec_separation}). FSI
uncertainties are assessed through comparisons among alternative FSI
implementations across event generators.

\begin{table}
  \caption{Main sources of systematic uncertainty affecting $\RR$ and
    their expected cancellation behavior in the ratio. Full
    cancellation indicates correlated effects between NC and CC
    samples. Partial indicates residual asymmetry. None indicates a
    source unique to one sample.}
\label{tab:systematics}
\begin{ruledtabular}
\begin{tabular}{l c}
Source & Cancellation \\
\hline
Flux normalization         & Full    \\
Flux shape                 & Partial \\
Proton reco.\ efficiency   & Full    \\
Calorimetric energy scale  & Full    \\
Proton threshold modeling  & Full    \\
Muon identification        & None    \\
Pion background            & Partial \\
  FSI model dependence
  \footnote{In the SRC-enhanced region, FSI contributes asymmetrically
  to NC (as the dominant production mechanism) and CC (as a smearing
  correction), so the degree of cancellation at large $\prel$ differs
  from that at low $\prel$. This asymmetry must be assessed through
  comparisons among different event generators.}       & Partial \\
MEC model dependence       & Partial \\
Nuclear ground state       & Partial \\
\end{tabular}
\end{ruledtabular}
\end{table}

\section{Discussion}
\label{sec:discussion}

\subsection{Separating SRC from meson-exchange currents}
\label{subsec:mec_separation}

Since MEC can reach the same region of pair phase space as SRC breakup
(Sec.~\ref{subsec:mec}), we summarize here the strategies that
separate the two mechanisms within this measurement.

The first and principal one is the ratio shape itself. MEC
contributes to both channels, so its normalization (and any
normalization uncertainty, experimental or theoretical) cancels in the
shape observable $\DR(\prel)$ of Eq.~(\ref{eq:double_ratio}). An
MEC-induced imitation of the SRC signal would require the isospin
composition of the emitted pairs to depend on the pair kinematics
asymmetrically between the NC and CC channels [cf.\
Eq.~(\ref{eq:mech_decomposition})]. This is a well-defined model
prediction, dominated by the $\Delta$
current~\cite{RuizSimo:2016pairs}, that can be evaluated with the
exclusive 2p2h calculations now available~\cite{Sobczyk:2020dhh,
Kasturi:2026exc} and bounded with CC~$0\pi$
data~\cite{Abratenko:2022nmc}.

A second strategy uses only the two proton momenta and is therefore
available in the NC as well as the CC sample: the momentum sharing
within the pair, quantified by the asymmetry $\alpha = (|\bm{p}_1| -
|\bm{p}_2|)/(|\bm{p}_1| + |\bm{p}_2|)$. In SRC breakup the struck
nucleon absorbs the full momentum transfer while the spectator
retains a momentum of order $\krel$, so the pair is kinematically
asymmetric whatever the multinucleon dynamics. The sharing in 2p2h
emission is instead prescription dependent. Event generators
distribute the transferred energy--momentum democratically, decaying
the pair back to back in its center of mass, which yields symmetric
sharing on average. The exclusive calculation of
Ref.~\cite{Kasturi:2026exc} finds the opposite: the nucleon attached
to the weak vertex carries the dominant share, and the asymmetry
survives intranuclear rescattering. Binning $\RR$ in $\alpha$, which
requires no neutrino-energy or momentum-transfer reconstruction,
therefore serves two purposes. The $\alpha$ distributions of the
two-proton samples measure the sharing dynamics of the multinucleon
background, discriminating between the democratic and exclusive
treatments in the same data set. And the symmetric-sharing bins are
depleted of SRC breakup under either treatment, because the SRC
asymmetry is fixed by the spectator kinematics rather than by the
sharing model. These bins therefore measure the background asymmetry
$r_{\mathrm{bg}}(\prel)$ of Eq.~(\ref{eq:src_fraction}) in situ,
whatever mixture of multinucleon and rescattering events populates
them. Residual SRC leakage into them, concentrated at low momentum
transfer where the asymmetry is kinematically compressed, lowers the
reference and makes the extraction conservative, as in
Sec.~\ref{sec:observable}. A fall of $\RR$ with $\prel$ that is
present in the asymmetric-sharing bins but absent in the symmetric
ones would constitute the SRC signature with the background
hypothesis checked in parallel.

Third, the CC sample carries the muon, so
transverse-kinematic-imbalance variables~\cite{Lu:2015hea,
Dolan:2018sbb} can be built for the two-proton system and used to
characterize the mechanism composition of the denominator
independently of the ratio. Finally, antineutrino running flips the
isospin selection of both SRC and MEC production
(Sec.~\ref{subsec:extensions}), overconstraining the decomposition.

Even in the least favorable scenario, in which an MEC $\prel$ shape
partially mimics the SRC suppression, the measurement remains
useful. Inclusive CC~$0\pi$ data constrain the total
multinucleon strength but carry no information on its isospin
decomposition, which in current models is a pure theoretical
prediction of the two-body currents. No measurement of it exists in the
weak sector. Because this decomposition differs between neutrinos and
antineutrinos, it propagates directly into the $\nu/\bar\nu$
comparison from which $\delta_{\mathrm{CP}}$ is extracted. The two
possible outcomes of the measurement are therefore both informative.
A suppression of $\RR(\prel)$ exceeding the MEC-constrained
expectation establishes the SRC enhancement of the charged-current
channel. A suppression consistent with it constitutes a data-driven
constraint on the isospin structure of multinucleon emission, read
through the same charge-exchange correction. The depth of the 
suppression distinguishes further between the two
cases. By the counting of Eq.~(\ref{eq:isospin_floor}), the asymptote
of $\RR$ at large $\prel$ measures the $np$/$pp$ ratio of whatever mechanism dominates
the tail. SRC dominance predicts an approach toward the floor set by
the measured $C_{np}/C_{pp} \approx 18$--$20$, i.e.\
$\RR \to 0.015$--$0.017$ before FSI repopulation, whereas a
multinucleon-driven tail would settle at the level set by the isospin
structure of the two-body currents, a different and currently
unmeasured number. The asymptotic value is thus an isospin
measurement in its own right, rather than a consistency check.

\subsection{Future Extensions}
\label{subsec:extensions}

\textit{Antineutrino mode}: In $\numubar$ CC interactions,
$\numubar + p \to \mu^+ + n$ selects the proton member of a pair, so
the primary-vertex two-proton channel closes: striking the proton of
an $np$ pair leaves $nn$, and of a $pp$ pair leaves $np$. The
antineutrino double-proton ratio is therefore a background-dominated
control of the neutrino-mode measurement, and with neutron tagging
(below) the $\mu^+ np$ final state tags $pp$-SRC pairs, complementing
the neutrino-mode constraint on the $np$ sector.

\textit{Neutron tagging}: Experiments with future neutron-detection
capabilities could extend the observable to $pn$ final states,
enabling direct access to the $np$-SRC sector in NC interactions and
providing an overconstrained set of isospin measurements.

\textit{Nuclear target dependence}: Measuring $\RR$ on carbon, oxygen,
argon, and iron targets would test the $A$-dependence of the SRC
enhancement and probe the universality of the $\mathcal{R}_{np/pp}$
ratio across nuclei, connecting to the extensive JLab
database~\cite{Hen:2016kwp}.

\textit{Oscillation systematic constraints}: A precision measurement
of $\RR$ at the DUNE near detector would directly constrain the
leading hadronic systematic in NC/CC-ratio-based oscillation analyses,
benefiting standard $\delta_\mathrm{CP}$ measurements.

\section{Conclusions}
\label{sec:conclusions}

We have introduced $\RR(\prel,\pcm)$, the differential ratio of
neutral-current to charged-current double-proton production in
neutrino--argon scattering, as a probe of SRC-enhanced charged-current
dynamics.

\begin{itemize}

\item \textit{Mechanism.} Charged-current interactions convert the
dominant $np$-SRC pairs into visible two-proton final states, whereas
direct SRC breakup into two protons is isospin suppressed in the
neutral current. The suppression of $\RR$ at large $\prel$ therefore
reflects an enhancement of the CC denominator rather than a depletion
of the NC numerator.

\item \textit{Kinematics.} The momentum transfer dominates the
reconstructed $\pcm$, so the low-$\pcm$ SRC tag familiar from electron
scattering does not survive and the signal appears in $\prel$, where
in NuWro the suppression scales with the generator SRC fraction. The
practical observable is the projection $\RR(\prel)$, with $\pcm$
retained as a consistency check.

\item \textit{Robustness against MEC.} Multinucleon emission reaches
the same pair kinematics but contributes to both channels. It sets the
overall level of $\RR$, whereas in the simulation the fall with
$\prel$ appears only when the SRC mechanism is active. The shape
observable $\DR(\prel)$ removes this normalization, together with
muon-identification and overall MEC and FSI uncertainties. Within the
two-component picture, the measured $\RR(\prel)$ determines
the SRC fraction of the CC two-proton sample, from a background reference
measured in situ and an SRC reference value bounded from below by the isospin
floor. Even if part of the
shape were found to arise from multinucleon dynamics, the measurement
would still provide the first constraint on the isospin composition of
weak two-nucleon emission.

\item \textit{Systematics and feasibility.} The ratio cancels the
flux normalization and strongly suppresses correlated detector
systematics, while the normalized observable $\DR(\prel)$ further
removes the dominant normalization uncertainties associated with muon
identification, MEC, and FSI. At SBND exposures the measurement is not
statistics limited, making the SRC-induced suppression experimentally
accessible with existing liquid-argon time projection chambers.


\end{itemize}

We encourage the experimental collaborations to carry out this
measurement as a direct test of the predicted SRC enhancement of
charged-current two-proton production.

\begin{acknowledgments}
The authors thank Ornella Palamara for stimulating discussions on
nuclear effects in neutrino scattering, which were the seed for this
project. We thank Rwik Dharmapal and Hemant Prasad for help with
the NuWro simulations. This work is supported by the Spanish Ministry
of Science and Innovation under grant PID2023-147949NB-C53.
\end{acknowledgments}

\bibliographystyle{apsrev4-2}
\bibliography{references}

\end{document}